\documentclass[10pt,a4paper]{article}

\usepackage{times}
\usepackage{comment}

\usepackage{epsfig,wrapfig}
\usepackage{epstopdf} 

\usepackage{fancyhdr}

\usepackage[lmargin=2.5cm, rmargin=2.5cm,tmargin=3.50cm,bmargin=3.50cm]{geometry}
\usepackage{indentfirst}

\usepackage{titlesec}
\titleformat{\section}[hang]
  {\centering}{\thesection}{1ex}{\normalsize \textsc}
\titleformat{\subsection}[hang]
  {}{\thesubsection}{1ex}{\normalsize \textit}

\usepackage{graphicx} 
\usepackage{amsmath} 
\usepackage{amssymb} 
\usepackage{braket}
\usepackage{physics} 
\usepackage{xcolor}
\usepackage{mathtools}
\usepackage{xcolor}

\newcommand{\pati}[1]{}

\renewcommand{\thesection}{ \normalsize \textnormal{\Roman{section}.}}
\renewcommand{\thesubsection}{\normalsize \textnormal{\textsc{\textit{\Alph{subsection}.}}}}

\def\e{\begin{equation}}
\def\f{\end{equation}}
\def\_#1{{\bf #1}}

\def\.{\cdot}

\renewcommand{\eqref}[1]{(\ref{#1})}

\begin{document}

\title{\vspace{-9mm}\large \textbf{Photon Transitions in Arbitrary Time-Varying Metamaterials}\vspace{-5mm}}

\vspace{-6ex}
\def\affil#1{\begin{itemize} \item[] #1 \end{itemize}}
\author{\vspace{-5cm} \normalsize \bfseries \underline{A. Stevens}$^1$ and C. Caloz$^1$}
\date{}
\maketitle
\thispagestyle{fancy} 
\vspace{-6ex}
\affil{\begin{center}\normalsize $^1$KU Leuven, Department of Electrical Engineering, Kasteelpark Arenberg, 3001, Leuven, Belgium \\
artuurstevens@hotmail.com
\vspace{-3mm}
 \end{center}}

\begin{abstract}
\noindent \normalsize
\textbf{\textit{Abstract} \ \ -- \ \
We present a general theory for calculating photon transitions in arbitrarily time-varying metamaterials. This theory circumvents the difficulties of conventional approaches in solving such a general problem by exploiting the eigenstates of time-dependent number operators. We demonstrate here the temporal evolution of these operators and the related transition probabilities for the cases of logistic and linear permittivity profiles. The theory is potentially extensible to arbitrary space-time modulations and may hence lead to multiple novel quantum effects and applications.}
\end{abstract}

\section{Introduction}
\vspace{-3mm}

\pati{GSTEMs}

Generalized Space-Time Engineered Modulation metamaterials, or GSTEMs for short, are structures that exploit the modulation of some parameters of a host medium in space and time for unprecedented opportunities in manipulating waves~\cite{Caloz_APM_12_2022}. Although relatively recent, they have already led to many novel concepts and applications, including parametric amplification~\cite{Holberg_1966}, the inverse prism~\cite{Akbarzadeh_OL_2018}, nonreciprocal devices~\cite{Chamanara_2017}, temporal aiming~\cite{Pacheco_2020}, new gain mechanisms~\cite{Pendry_2021}, arbitrary chirping~\cite{Kinder_2024} and gravity analogs~\cite{Bahrami_2023}.

\pati{Related Works}
The vast majority of studies on GSTEMs to date have concerned the classical regime and very little work has been on GSTEMs in the \emph{quantum} regime. A few exceptions, all restricted to purely temporal modulations~\cite{Morgenthaler_1958}, include the following. Mendonça~\textit{et al.} investigated photon production and annihilation at an abrupt permittivity discontinuity~\cite{Mendonça_2000}. Mirmoosa~\textit{et al.} extended that work to include correlation functions~\cite{Mirmoosa_2023}. Ganfornina-Andrades~\textit{et al.} examined energy production from vacuum in a arbitrary permittivity-permeability modulation~\cite{Liberal_2023}. However, these reports considered transition probabilities only for an abrupt modulation~\cite{Mendonça_2000,Mirmoosa_2023} or arbitrary modulations without considering transitions probabilities~\cite{Liberal_2023}.

\pati{Contribution}
Here, we generalize the works mentioned above by calculating the transition probabilities for arbitrary time modulations. We show that the Schrödinger equation and time-evolution operator approaches entail major difficulties in such a general system. Therefore, we present an alternative in the Heisenberg picture using eigenstates of the time-dependent number operators. Based on these eigenstates, we derive a closed-form solution to the problem and show related results when the initial state is vacuum.
\section{Hamiltonian}\label{sec:Heisenberg}
\vspace{-3mm}
\pati{Hamiltonian}
The Hamiltonian of our problem, assuming no charges or currents, reads $ \hat{H} = \int_V d^3 r \left( \frac{\hat{\mathbf{D}}^2}{2\epsilon(t)} + \frac{\hat{\mathbf{B}}^2}{2\mu(t)}\right) $, where $\epsilon(t)$ and $\mu(t)$ are the time-varying permittivity and permeability of the medium \cite{Hillery_2009}. Expanding $\hat{\mathbf{D}}$ and $\hat{\mathbf{B}}$ in the creation and annihilation operators, $\hat{a}^\dagger_{\textbf{k}, \lambda}(t)$ and $\hat{a}_{\textbf{k}, \lambda}(t)$, leads to 
\begin{equation}\label{eq:H}
\hat{H}(t) =\frac{i\hbar}{2} \sum_{\textbf{k}, \lambda}  \alpha_k(t) \left(   2\hat{n}_{\textbf{k}, \lambda}(t) +1     \right) +   \beta_k(t)  \left(\hat{a}_{\textbf{k}, \lambda}(t)  \hat{a}_{-\textbf{k}, \lambda}(t)  +
 \hat{a}^\dagger_{\textbf{k}, \lambda}(t) \hat{a}^\dagger_{-\textbf{k}, \lambda}(t)                    \right),
\end{equation}
where $\alpha_k(t) = \frac{-i\omega_k}{2}\left( \frac{\epsilon(0)}{\epsilon(t)} + \frac{\mu(0)}{\mu(t)}  \right)$ and 
$\beta_k(t) = \frac{(-1)^{\lambda + 1}i\omega_k}{2}\left( \frac{\epsilon(0)}{\epsilon(t)} - \frac{\mu(0)}{\mu(t)}  \right)$. Before the onset ($t=0$) of the modulation ($\epsilon(t<0),\mu(t<0)$ constant), we have $\beta_k(t) = 0$, so that $\hat{H} $ reduces to $\hat{H} =\sum_{\mathbf{k},\lambda} \hbar \omega_{k} \left(\hat{n}_{\mathbf{k}, \lambda}(t) + 1/2\right)$~\cite{Hillery_2009}-\cite{Loudon_2000}. Here, $\hat{n}_{\mathbf{k}, \lambda}(t)=\hat{a}^\dagger_{\textbf{k}, \lambda}(t) \hat{a}_{\textbf{k}, \lambda}(t)$, which is also present in~Eq.~\eqref{eq:H}, is the number operator, whose eigenstates for $t<0$, $\ket{n}$, represent states with $n$ photons (Fock states) in the mode $(\mathbf{k}, \lambda)$. Thus, the term $\hat{a}_{\textbf{k}, \lambda}(t)  \hat{a}_{-\textbf{k}, \lambda}(t)$ in Eq.~\eqref{eq:H} represents the annihilation of a photon in the forward mode ($\mathbf{k}$) with a photon in the backward mode ($-\mathbf{k}$), while the term $\hat{a}^\dagger_{\textbf{k}, \lambda}(t) \hat{a}^\dagger_{-\textbf{k}, \lambda}(t)$ represents the emission of two photons propagating in opposite directions. These processes correspond to photon transitions between Fock states as the system evolves in time.

\pati{Heisenberg Picture}
In order to calculate the probabilities of these transitions, we have to determine the time-varying number operator. This may be accomplished by solving the Heisenberg equation, $i\hbar\dot{\hat{a}}_{\textbf{k},\lambda}(t)= [\hat{a}_{\textbf{k},\lambda}(t),\hat{H}]$, with Eq.~\eqref{eq:H}, viz., 
\begin{equation}\label{diff_eq}
    \frac{d\hat{a}_{\textbf{k},\lambda}(t)}{dt} = \alpha_k(t) \hat{a}_{\textbf{k}, \lambda}(t) + \beta_k(t) \hat{a}^\dagger_{-\textbf{k}, \lambda}(t).
\end{equation}
The form of Eq.~\eqref{diff_eq} suggests the ansatz 
\begin{equation}\label{eq:ansatz}
    \hat{a}_{\textbf{k},\lambda}(t)=f_k(t)\hat{a}_{\textbf{k}, \lambda}(0) + g_k(t)\hat{a}^\dagger_{-\textbf{k}, \lambda}(0).
\end{equation}
Inserting this relation into Eq.~\eqref{diff_eq}, and separating the terms in $\hat{a}_{\textbf{k},\lambda}(0)$ and $\hat{a}^\dagger_{-\textbf{k},\lambda}(0)$, yields
\begin{equation}\label{diff_eq_f}
    \frac{df_k}{dt} = \alpha_k(t)f_k(t) +  \beta_k(t)g_k^*(t)
    \quad\text{and}\quad
    \frac{dg_k}{dt} = \alpha_k(t)g_k(t) +  \beta_k(t)f_k^*(t).
\end{equation}
Once $\hat{a}_{\textbf{k},\lambda}(t)$ has been determined upon solving Eqs.~\eqref{diff_eq_f}, $\hat{a}^\dagger_{\textbf{k},\lambda}(t)$ follows by hermitian conjugation, while $\hat{a}_{-\textbf{k},\lambda}(t)$ and $\hat{a}^\dagger_{-\textbf{k},\lambda}(t)$ follow by substituting $\textbf{k}$ by $-\textbf{k}$. Then, the number operator, $\hat{n}_{\mathbf{k}, \lambda}(t)=\hat{a}^\dagger_{\textbf{k}, \lambda}(t) \hat{a}_{\textbf{k}, \lambda}(t)$, is fully determined.

\section{Transition Probabilities}
\vspace{-3mm}
\pati{Transition Probability}
Now we have to establish a connection between that number operator and the sought after transition probabilities. The transition probability to a state $\ket{u, v}$, denoting $u$ forward photons $(\textbf{k})$ and $v$ backward photons $(-\textbf{k})$, is then calculated as P$( \ket{u, v})(t) = |\bra{u,v} \ket{\psi(t)}|^2$~\cite{Loudon_2000}, where the state $\ket{\psi(t)}$ evolves from its initial state $\ket{\psi(0)}$ as $\ket{\psi(t)} = \hat{U}(t) \ket{\psi(0)}$, with $\hat{U}(t)$ being the time-evolution operator $\hat{U}(t) = T\left\{ \textrm{e}^{\frac{1}{i\hbar} \int_0^t d\tau \hat{H}(\tau)} \right\}$~\cite{Mandl}. Unfortunately, the calculation of $\hat{U}(t)$, involving, upon Taylor expansion of the exponential function, a series of time-ordered integrals, is extremely complicated, a complexity that is inherent to $\hat{U}(t)$. One may then perhaps bypass that operator by combining it into a block with the number state as $\hat{U}^\dagger(t)\ket{u,v}$, which is an eigenstate of the time-dependent number operators $\hat{n}_{\mathbf{k}, \lambda}(t)$ and $\hat{n}_{-\mathbf{k}, \lambda}(t)$, with eigenvalues $u$ and $v$, respectively. One could also attempt to solve the Schrödinger equation, but this leads to an infinite set of coupled differential equations which are very complex to solve as well.

\pati{Time-dependent Eigenstates}
Let us then rewrite the transition probability as $\textrm{P}( \ket{u, v})(t) = |\bra{\psi(0)} \hat{U}^\dagger(t) \ket{u,v} |^2$ by taking the hermitian conjugate of the previous formula. Using the shorthand notation $\ket{\xi_{u,v}(t)} = \hat{U}^\dagger(t)\ket{u,v}$, and recursively applying the creation operators $\hat{a}^\dagger_{\textbf{k}, \lambda}(t)\ket{\xi_{u,v}(t)} = \sqrt{u+1}e^{i\phi_u} \ket{\xi_{u+1,v}(t)}$ and $\hat{a}^\dagger_{-\textbf{k}, \lambda}(t)\ket{\xi_{u,v}(t)} = \sqrt{v+1}e^{i\phi_v} \ket{\xi_{u,v+1}(t)}$, yields then
\begin{equation}\label{TD Fock states}
\begin{split}
\ket{\xi_{u,v}(t)} &= \frac{\left(\hat{a}^\dagger_{\textbf{k}, \lambda}(t) \right)^u \left(\hat{a}^\dagger_{-\textbf{k}, \lambda}(t) \right)^v}{\sqrt{u!v!}}e^{-i \Phi_{u,v}(t)}  \ket{\xi_{0,0}(t)},
\end{split}
\end{equation}
where $e^{-i \Phi_{u,v}(t)}$ is a phase factor that does not affect the transition probability and can hence be ignored. Moreover, the ground state $\ket{\xi_{0,0}(t)}$ may be found by expansion in Fock states, $\ket{\xi_{0,0}(t)} = \sum_{n, m} C_{n, m} \ket{n, m}$, and application of the destruction rules $\hat{a}_{\textbf{k}, \lambda}(t)\ket{\xi_{0,0}(t)} = 0$ and $\hat{a}_{-\textbf{k}, \lambda}(t)\ket{\xi_{0,0}(t)} = 0$; this yields an expression for $C_{n,m}$ in terms of $f_k(t)$ and $g_k(t)$, and hence determines $\ket{\xi_{0,0}(t)}$ in terms of these functions. Then, using also the construction operators established in Sec.~\ref{sec:Heisenberg}, the state $\ket{\xi_{u,v}(t)}$ is completely determined, and a closed-form of the transition probability formula can then be obtained upon inserting $\ket{\psi(0)}$ into $\textrm{P}( \ket{u, v})(t) = |\bra{\psi(0)} \ket{\xi_{u,v}(t)} |^2$.

\pati{Examples}
We shall consider here only the case where the initial state is the vacuum state, i.e., $\ket{\psi(0)} = \ket{0, 0}$. The transition formula takes then the simple form
\begin{equation}\label{eq:trans_prob}
\textrm{P}(\ket{u,v})(t)  =\frac{1}{1+|g_k(t)|^2} \left( \frac{|g_k(t)|^{2}}{1+|g_k(t)|^2}   \right)^u  \delta_{u, v},
\end{equation}
which is seen to depend only on the coupling function $g_k(t)$ and not on the intrinsic function $f_k(t)$. Interestingly, as revealed by the delta function, the only states to which vacuum can evolve are states with the same number of forward and backward photons, which automatically ensures the conservation of momentum. The formula~\eqref{eq:trans_prob} also reveals that at any time where $g_k(t)$ vanishes (using $0^0=1$), the system returns to the vacuum state.

\begin{figure}[h]
    \centering
    \includegraphics[width = \linewidth]{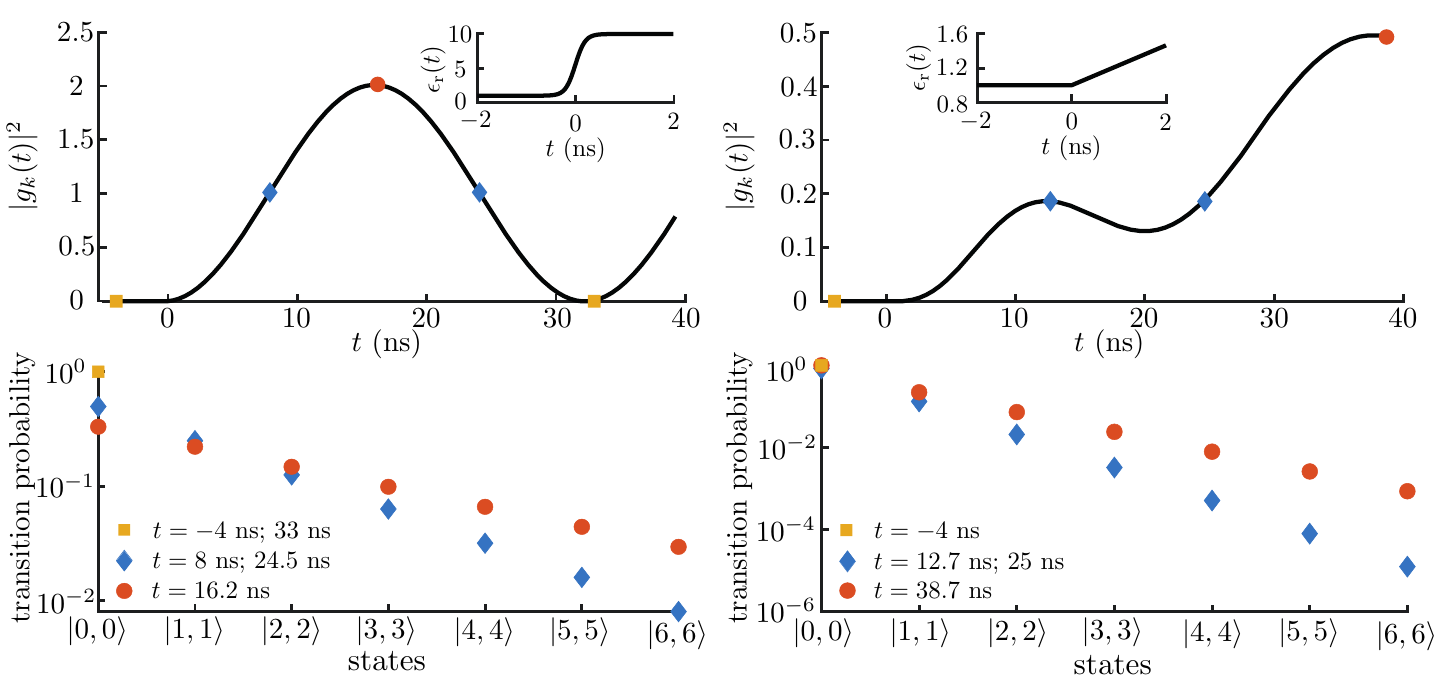}
    \vspace{-8mm}
    \caption{Time evolution of the coupling function $|g_k(t)|^2$ [Eq.~\eqref{eq:ansatz}] (top) and of the transition probabilities [Eq.~\eqref{eq:trans_prob}] (bottom) for two different time-varying permittivity (and constant permeability) media and $k=10^7\textrm{m}^{-1}$. (Left)~Smooth (logistic function) permittivity profile. (Right)~Linear permittivity profile. }
    \label{prob}
\end{figure}

Figure~\ref{prob} shows the time evolution of the coupling function and transition probabilities for two different permittivity\footnote{The permeability function is kept constant for simplicity.} modulation profiles, a logistic step and linear increase. As expected, the system returns to the vacuum state at each time where the coupling function goes to zero. Moreover, the distribution function drops off slower when the coupling function is greater. Finally, the shape of the modulation function (insets) affects the coupling function. In the logistic case, the coupling function varies periodically (in fact, sinusoidally), while in the linear case it grows on average monotonically. The former case corresponds to a sort of global Rabi oscillations, which requires further investigation, while the latter case is essentially understood from the fact that the amount of energy provided by the modulation constantly grows.

\end{document}